\documentclass[english,aps,superscriptaddress,preprintnumbers,reprint,footinbib,amsmath,amssymb,prbr]{revtex4-1}
\usepackage[latin9]{inputenc}
\usepackage{amsmath}
\usepackage{amssymb}
\usepackage{graphicx}
\usepackage{esint}

\makeatletter

\newcommand*\LyXThinSpace{\,\hspace{0pt}}

\@ifundefined{textcolor}{}{%
 \definecolor{BLACK}{gray}{0}
 \definecolor{WHITE}{gray}{1}
 \definecolor{RED}{rgb}{1,0,0}
 \definecolor{GREEN}{rgb}{0,1,0}
 \definecolor{BLUE}{rgb}{0,0,1}
 \definecolor{CYAN}{cmyk}{1,0,0,0}
 \definecolor{MAGENTA}{cmyk}{0,1,0,0}
 \definecolor{YELLOW}{cmyk}{0,0,1,0}
}

\usepackage{ulem}

\usepackage{aecompl}

\usepackage{epsfig}\usepackage{dcolumn}\usepackage{bm}

\usepackage{babel}

\makeatother

\usepackage{babel}
\begin{document}
\title{Atomic frustrated impurity states in Weyl metals}
\author{W. N. Mizobata}
\affiliation{São Paulo State University (Unesp), School of Engineering, Department
of Physics and Chemistry, 15385-000, Ilha Solteira-SP, Brazil}
\author{Y. Marques}
\affiliation{Department of Physics, ITMO University, St.~Petersburg 197101, Russia}
\author{M. Penha}
\affiliation{São Paulo State University (Unesp), School of Engineering, Department
of Physics and Chemistry, 15385-000, Ilha Solteira-SP, Brazil}
\author{J. E. Sanches}
\affiliation{São Paulo State University (Unesp), School of Engineering, Department
of Physics and Chemistry, 15385-000, Ilha Solteira-SP, Brazil}
\author{L. S. Ricco}
\affiliation{São Paulo State University (Unesp), School of Engineering, Department
of Physics and Chemistry, 15385-000, Ilha Solteira-SP, Brazil}
\author{M. de Souza}
\affiliation{São Paulo State University (Unesp), IGCE, Department of Physics, 13506-970,
Rio Claro-SP, Brazil}
\author{I. A. Shelykh}
\affiliation{Department of Physics, ITMO University, St.~Petersburg 197101, Russia}
\affiliation{Science Institute, University of Iceland, Dunhagi-3, IS-107, Reykjavik,
Iceland}
\author{A. C. Seridonio}
\email[corresponding author: ]{antonio.seridonio@unesp.br}

\affiliation{São Paulo State University (Unesp), School of Engineering, Department
of Physics and Chemistry, 15385-000, Ilha Solteira-SP, Brazil}
\affiliation{São Paulo State University (Unesp), IGCE, Department of Physics, 13506-970,
Rio Claro-SP, Brazil}
\begin{abstract}
We theoretically analyze the effect of the inversion symmetry breaking
on the structure of the impurity molecular states in Weyl metals.
We show that for the case of a highly noncentrosymmetric Weyl metallic
host, the standard picture of the alternating bonding and antibonding
orbitals breaks down, and a qualitatively different frustrated atomic
state emerges. This is a consequence of the pseudogap closing and
related delicate Fano interplay between intra- and inter-impurity
scattering channels.
\end{abstract}
\maketitle

\section{Introduction}

Dirac-Weyl equation~\cite{Weyl1929}, which first appears in the
context of the relativistic quantum field theory, where it describes
massless fermions, such as neutrinos, recently found its application
in the domain of condensed matter physics. The existence of Dirac-Weyl
fermions, quasi-relativistic quasiparticles, was unambiguously demonstrated
for the family of the gapless binary alloys, such as Na$_{3}$Bi,
Cd$_{3}$As$_{2}$, TaAs, NbAs and TaP \cite{Wang2012,Liu2014a,Wang2013,Liu2014b,Huang2015,Weng2015,Xu2015I,Lv2015I,Lv2015II,Xu2015II,Xu2016}.
The pair of the Dirac cones, present in these materials, can be split
into two Weyl nodes with opposite chirality, if certain symmetry (inversion
or time-reversal) is broken \cite{Armitage2018}. As a result, a topological
Weyl material with unusual characteristics, such as Fermi arcs, chiral
anomaly and exotic Hall effects \cite{Armitage2018,Wan2011,Yang2011,Hosur2012,Kim2017,Nielsen1983,GXu2011},
emerges. The peculiar band structure of Weyl systems has dramatic
impact on the electronic structure of impurities \cite{Sun2015,Ma2015,Chang2015,Hosseini2015,Principi2015,Zheng2016,Marques2017,Marques2019}.
In particular, as it was recently shown by some of us, chiral magnetic
chemical bounds for a pair of impurities can appear in Weyl semimetals
with energy degenerate Weyl nodes shifted in $\textbf{k}$ space with
respect to each other \cite{Marques2019}.

\begin{figure}[!]
\centering\includegraphics[width=0.45\textwidth]{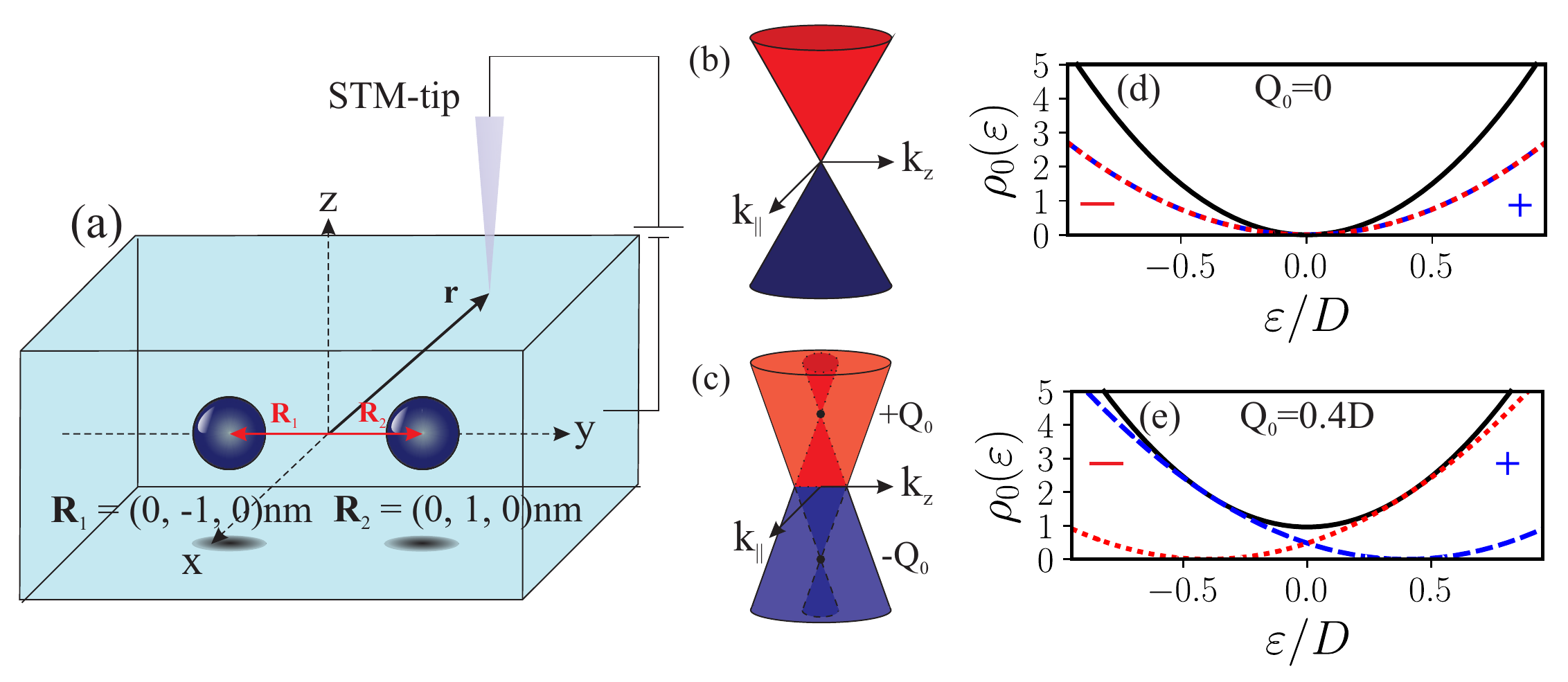}\caption{\label{fig:Pic1} {(Color online) Panel (a): Sketch of the considered
system, consisting of a pair of impurities placed inside a Weyl metal
close to its interface. The positions of the impurities are characterized
by the vectors $\textbf{R}_{1,2}$. The impurity molecular states
can be probed on the surface of the host by an STM tip, whose location
is characterized by the vector $\textbf{r}$. Panel (b): Sketch of
the dispersion, characteristic for a Dirac semimetal with two degenerated
Dirac cones. The pseudogap is formed around the Dirac point, where
the host Density of States (DOS) $\rho(\varepsilon)=0$. Panel (c):
Sketch of the dispersion, characteristic for the Weyl metal. The degeneracy
of the Dirac cones is lifted due to the breaking of the inversion
symmetry, and a pair of Weyl nodes vertically shifted with respect
to each other appears. The pseudogap is closed due to the lifting
of the degeneracy of the Weyl nodes. Panel (d): DOS $\rho(\varepsilon)$
of a Dirac semimetal. Panel (e): DOS $\rho(\varepsilon)$ of a Weyl
metal. The plus and minus signs identify the DOS resolved in opposite
chiralities.}}
\end{figure}

In this communication, we consider the structure of impurity molecular
states in Weyl metals, where two Weyl nodes are located at the same
${\bf k}$, but are shifted in energy. We demonstrate that in the
geometry corresponding to two Anderson-like impurities \cite{Anderson1961}
shown in Fig.~\ref{fig:Pic1}, bonding and antibonding molecular
states evolve into an atomic frustrated state marked by two Hubbard
bands \cite{Hubbard1963}, with increase of the energy splitting between
the two Weyl nodes. In this regime, the closing of the host pseudogap
occurs, which leads to the dominance of the destructive Fano interference
\cite{Fano,FanoReview} in the intra-impurity scattering channel,
which is opposite to what happens in the corresponding inter-impurity
channel revealing resonant behavior. The reported crossover can be
realized by application of external stress \cite{Armitage2018} and
experimentally detected with use of the STM techniques.

\section{The Model }

The Hamiltonian of the system sketched in Fig.~\ref{fig:Pic1} can
be represented as:

\begin{align}
\mathcal{H} & =\sum_{\mathbf{k}}\psi^{\dagger}(\mathbf{k})(H_{+}\oplus H_{-})\psi(\mathbf{k})+\varepsilon_{d}\sum_{j\sigma}d_{j\sigma}^{\dagger}d_{j\sigma}\nonumber \\
 & +U\sum_{j}d_{j\uparrow}^{\dagger}d_{j\uparrow}d_{j\downarrow}^{\dagger}d_{j\downarrow}+\sum_{j\mathbf{k}}\tilde{d}_{j}^{\dagger}\tilde{V}_{j\mathbf{k}}\psi(\mathbf{k})+\text{H.c.},\label{eq:Hamiltonian}
\end{align}
where $H_{\chi}(\mathbf{k})=\chi(v_{F}\boldsymbol{\sigma}\cdot\mathbf{k}+\sigma_{0}Q_{0})$
is the Dirac-Weyl Hamiltonian of the host, corresponding to the two
Dirac cones shifted vertically in energy (see Fig. \ref{fig:Pic1}(c)),
$\boldsymbol{\sigma}$ stands for the vector of Pauli matrices, $\sigma_{0}$
is the unity matrix, $\chi=\pm1$ corresponds to the Weyl nodes chirality,
$Q_{0}$ is the characteristic parameter defining the energy splitting
between the Weyl nodes ($Q_{0}\neq0$ corresponds to a Weyl metal,
$Q_{0}=0$ to a Dirac semimetal), $v_{F}$ is the Fermi velocity,
$\psi(\mathbf{k})=(c_{\mathbf{k}+\uparrow},c_{\mathbf{k}+\downarrow},c_{\mathbf{k}-\uparrow},c_{\mathbf{k}-\downarrow})^{T}$
is the four-spinor operator describing the electronic states in the
host $c_{\mathbf{k}\chi\sigma}^{\dagger},c_{\mathbf{k}\chi\sigma}$
with wave vector $\bold k$, chirality $\chi$ and spin $\sigma$.
The operators $d_{j\sigma}^{\dagger},d_{j\sigma}$ describe the electronic
states of individual impurities ($j=1,2$) with single-particle energies
$\varepsilon_{d}$ and on-site Coulomb correlation energy $U$. The
term, containing the two-spinor $\tilde{d}_{j}^{\dagger}=\begin{array}{cc}
(d_{j\uparrow}^{\dagger},d_{j\downarrow}^{\dagger}),\end{array}$ couples the impurities to the host, via the matrix
\begin{eqnarray}
\tilde{V}_{j\mathbf{k}} & =v_{0} & \left(\begin{array}{cc}
e^{i\mathbf{k}\cdot\mathbf{R}_{j}} & 0\\
0 & e^{i\mathbf{k}\cdot\mathbf{R}_{j}}
\end{array}\begin{array}{cc}
e^{i\mathbf{k}\cdot\mathbf{R}_{j}} & 0\\
0 & e^{i\mathbf{k}\cdot\mathbf{R}_{j}}
\end{array}\right),\label{eq:Vkmatrix-1}
\end{eqnarray}
with $v_{0}$ being the coupling strength.

The electronic characteristics of the system are determined by its
Local Density of States (LDOS) $\rho(\varepsilon,{\bold r})$, which
can be found from the Green's functions (GF) of the host in the energy
domain, $\tilde{\mathcal{G}}_{\chi\chi'\sigma}(\varepsilon,{\bold r})$
\cite{ManyBody} defined as the time-Fourier transform of $\mathcal{G}_{\chi\chi'\sigma}(t,{\bold r})=-i\theta\left(t\right)\left\langle \{\psi_{\chi\sigma}(t,{\bold r}),\psi_{\chi'\sigma}^{\dagger}(0,{\bold r})\}\right\rangle _{\mathcal{H}},$
with $\psi_{\chi\sigma}(t,{\bold r})=\sum_{\mathbf{k}}e^{i\mathbf{k}\cdot\mathbf{r}}c_{\bold k\chi\sigma}(t)$
being the field operator of the host conduction states with spin $\sigma$
and chirality $\chi$. The LDOS reads \cite{Marques2017,Marques2019,ManyBody}:
\begin{equation}
\rho(\varepsilon,{\bold r})=-\frac{1}{\pi}\sum_{\sigma\chi\chi'}\textrm{Im}\{\tilde{\mathcal{G}}_{\chi\chi'\sigma}(\varepsilon,{\bold r})\}=\rho_{0}+\sum_{jj'}\delta\rho_{jj'},\label{eq:LDOS}
\end{equation}
where the first term in this expression describes the host DOS $\rho_{0}=\sum_{\chi}\frac{3\varepsilon_{\chi}^{2}}{D^{3}},$
with $D$ as the energy cutoff and $\varepsilon_{\chi}=\varepsilon-\chi Q_{0}$,
and the second term is the correction to the LDOS induced by the host-impurity
coupling:

\begin{align}
\delta\rho_{jj'}(\varepsilon,\bold r) & =-\frac{1}{\pi v_{0}^{2}}\sum_{\chi\chi'\sigma}\textrm{Im}[\Sigma_{\chi\sigma}^{+}\bigl({\bold r-\bold R_{j}}\bigr){\cal {\cal \tilde{G}}}_{j\sigma|j'\sigma}(\varepsilon)\nonumber \\
 & \times\Sigma_{\chi'\sigma}^{-}\bigl({\bold r-\bold R_{j'}}\bigr)],\label{eq:LDOSjl}
\end{align}
where $\bold R_{j}$ describes the coordinates of the two impurities.
The terms with $j'=j$ and $j'\neq{j}$ correspond to intra- and inter-impurity
scattering channels, respectively, and
\begin{eqnarray}
\Sigma_{\chi\sigma}^{\pm}({\bold r})=-\frac{3\pi v_{F}v_{0}^{2}}{2D^{3}}\frac{e^{-i|{\bold r}|\frac{\varepsilon_{\chi}}{v_{F}}}}{|{\bold r}|}\left[\varepsilon_{\chi}\pm\chi\sigma\left(\varepsilon_{\chi}+i\frac{v_{F}}{|{\bold r}|}\right)\right]\label{eq:SEr}
\end{eqnarray}
are self-energy terms responsible for the spatial modulation of the
LDOS. Following Ref.\cite{Zheng2016}, to obtain Eq.(\ref{eq:SEr}),
we have evaluated the noninteracting part of the GF $v_{0}^{2}\tilde{\mathcal{G}}_{\chi\chi'\sigma}(\varepsilon,{\bold r})$
(the corresponding one which solely considers the first term in Eq.(\ref{eq:Hamiltonian})),
by means of an expansion of the plane wave $e^{i\mathbf{k}\cdot{\bold r}}$
within $\tilde{\psi}_{\chi\sigma}(\varepsilon,{\bold r})$ (the Fourier
transform of $\psi_{\chi\sigma}(t,{\bold r}))$ into spherical harmonics
terms, according to the Rayleigh equation, particularly for $D\gg\varepsilon$
\cite{Hosseini2015,Zheng2016}.

$\tilde{\mathcal{G}}_{j\sigma|j'\sigma}(\varepsilon)$ is the time-Fourier
transform of the impurities GFs, $\mathcal{G}_{j\sigma|j'\sigma}(t)=-i\theta\left(t\right)\bigl\langle\{d_{j\sigma}\left(t\right),d_{j'\sigma}^{\dagger}\left(0\right)\}\bigr\rangle_{\mathcal{H}}.$
Away from the Kondo regime \cite{Kondo1}, Hubbard-I approximation
\cite{Marques2017,Marques2019,Hubbard1963,ManyBody} can be applied,
which gives:
\begin{eqnarray}
\tilde{\mathcal{G}}_{j\sigma|j\sigma}(\varepsilon) & = & \frac{\lambda_{j}^{\bar{\sigma}}}{g_{j\sigma|j\sigma}^{-1}(\varepsilon)-\lambda_{j}^{\bar{\sigma}}{{\Sigma}_{\sigma}^{+}({\bold R_{12}})g_{j'\sigma|j'\sigma}(\varepsilon)\lambda_{j'}^{\bar{\sigma}}{\Sigma}_{\sigma}^{-}({\bold R_{12}})}}.\nonumber \\
\label{eq:Gjj}
\end{eqnarray}
Here $\bar{\sigma}=-\sigma$, $j'\neq j$, ${\bold R_{12}}={\bold R_{1}}-{\bold R_{2}}$,
${\Sigma}_{\sigma}^{\pm}({\bold r})=\sum_{\chi}\Sigma_{\chi\sigma}^{\pm}({\bold r}),$

\begin{equation}
g_{j\sigma|j\sigma}(\varepsilon)=\frac{1}{\varepsilon-\varepsilon_{j\sigma}-{\Sigma}_{0}}
\end{equation}
is the single-impurity noninteracting GF,
\begin{equation}
{\Sigma}_{0}=\frac{3v_{0}^{2}}{2D^{3}}\sum_{\chi}\varepsilon_{\chi}^{2}\left(\text{ln}\left|\frac{D+\varepsilon_{\chi}}{{D-\varepsilon_{\chi}}}\right|-\frac{2D}{\varepsilon_{\chi}}-i\right)\label{eq:SE0}
\end{equation}
as the local self-energy,
\begin{equation}
\lambda_{j}^{\sigma}=1+\frac{U}{g_{j\bar{\sigma}|j\bar{\sigma}}^{-1}(\varepsilon)-U}\bigl\langle n_{j\sigma}\bigr\rangle\label{eq:Lambda}
\end{equation}
is the spectral weight and
\begin{equation}
\bigl\langle n_{j\sigma}\bigr\rangle=-\frac{1}{\pi}\int_{-\infty}^{+\infty}n_{_{F}}(\varepsilon)\textrm{Im}[\mathcal{\tilde{G}}_{j\sigma|j\sigma}\left(\varepsilon\right)]d\varepsilon\label{eq:Number}
\end{equation}
is the impurity occupation~\cite{Kondo2}. The crossed GF reads

\begin{eqnarray}
\tilde{{\cal G}}_{{j\sigma}|{{j'}\sigma}}\left(\varepsilon\right) & = & g_{j\sigma|j\sigma}(\varepsilon){\lambda_{j}^{\bar{\sigma}}{\Sigma}_{\sigma}^{\pm}({\bold R_{j{j'}}})}{\cal \tilde{{\cal G}}}_{{{j'}\sigma}|{{j'}\sigma}}\left(\varepsilon\right),\label{eq:Gjl}
\end{eqnarray}
in which the $\pm$ signs correspond to $j=1,j'=2$ and $j=2,j'=1$,
respectively. We emphasize that to close the set of Eqs. (\ref{eq:Gjj}),
(\ref{eq:Lambda}) and (\ref{eq:Gjl}) we have followed, as mentioned
previously, the Hubbard-I scheme \cite{Marques2017,Marques2019,Hubbard1963,ManyBody}.
It truncates the GFs of the Hamiltonian (Eq.(\ref{eq:Hamiltonian}))
by taking into account the Coulomb blockade regime \cite{Hubbard1963,ManyBody}
and neglecting the Kondo correlations, where spin-flip processes dominate
\cite{Kondo1}. Thus, in applying the equation of motion procedure
(EOM) to evaluate such GFs \cite{ManyBody}, those showing spin-flip
scattering should be disregarded. By this manner, the impurity occupation
$\bigl\langle n_{j\sigma}\bigr\rangle$ (Eq.(\ref{eq:Number})) is
then determined off the Kondo limit \cite{Kondo2}, by performing
a self-consistent calculation. The unique regime in which the equations
above are exact is for $U=0$.

In the case of uncorrelated impurities, realized when $|\bold R_{12}|\gg v_{F}v_{0}^{2}/D^{3}$,
${\Sigma}_{\sigma}^{\pm}({\bold R_{j{j'}}})=0$ and $\delta\rho_{jj'}=0$,
Eq.~(\ref{eq:Gjj}) has two poles (the so-called Hubbard resonant
bands~\cite{Hubbard1963}), appearing in $\delta\rho_{jj}$. The
host-mediated inter-impurity correlations lead to the splitting of
these poles, which corresponds to the formation of the impurity molecular
bands even in the absence of the direct hopping term between the impurities
\cite{Marques2017}.

\section{Results and Discussion}

In our following consideration, we use model parameters: $|{\bold R_{12}}|=2\:\text{{nm}}$,
$\varepsilon_{d}=-0.07D$, $v_{0}=-0.14D$, $U=0.14D$, $v_{F}\approx3\:\text{eV{Å}}$
and $D\approx0.2\,\text{\text{{eV}}}$ \cite{Marques2017,Marques2019}.
We suggest that the impurities are buried at the distance of $1\:\text{{nm}}$
below the top surface of the Dirac-Weyl material, and are placed in
the points $\textbf{R}_{1}=(0,-1,0)\:\text{{nm}}$ and $\textbf{R}_{2}=(0,1,0)\:\text{{nm}}$
(see Fig. \ref{fig:Pic1}).

\begin{figure}[!]
\centering\includegraphics[width=0.48\textwidth]{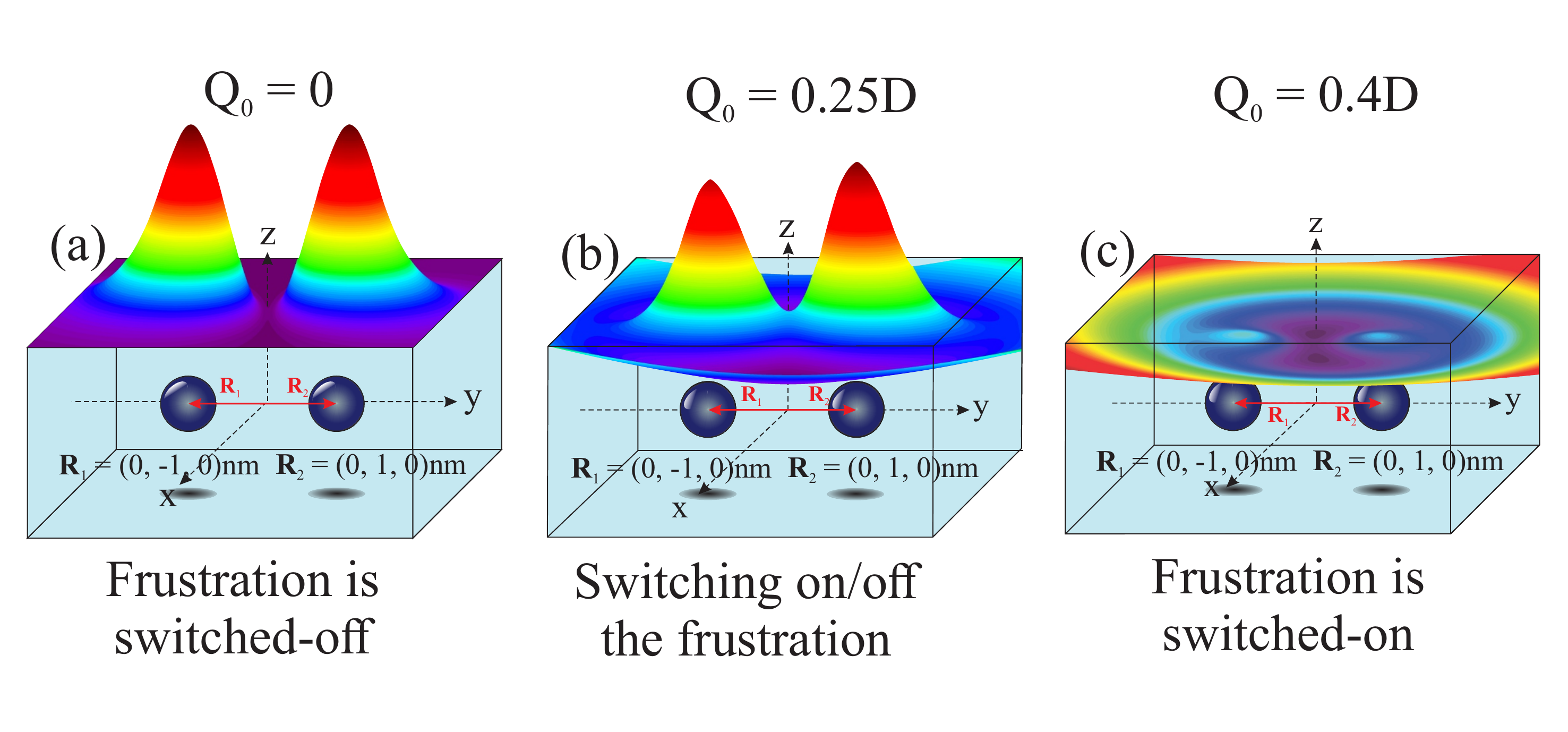}\caption{\label{fig:Pic2} {(Color online) Panel (a): Spatial profile of the
LDOS, corresponding to the antibonding state of a pair of impurities,
placed inside a Dirac semimetal ($Q_{0}=0$). Panel (b): Spatial profile
of the LDOS for a pair of impurities, placed inside a Weyl metal with
moderate value of $Q_{0}=0.25D$. Panel (c): Spatial profile of the
LDOS, corresponding to the frustrated atomic state, for a pair of
impurities, placed inside a Weyl metal with large value of $Q_{0}=0.4D$.}}
\end{figure}

Fig.~\ref{fig:Pic2} illustrates the evolution of the spatial profiles
of the LDOS at the surface of the host, given by Eq.~(\ref{eq:LDOS}),
which can be probed by an STM tip, with increase of the parameter
$Q_{0}$, describing the breaking of the inversion symmetry. In panel
(a) the case of a Dirac semimetal with degenerated Weyl nodes, corresponding
to $Q_{0}=0$, is illustrated. Molecular orbitals of the bonding and
antibonding type are formed, and the profile corresponding to the
latter one, with maxima of the LDOS centered at the points where the
impurities are located, is shown. We stress that due to the peculiarities
of the band structure of the Dirac host, the antibonding state has
lower energy as compared to the bonding state, as it was demonstrated
in Ref.~\cite{Marques2017}. The increase of the parameter $Q_{0}$
leads to the broadening of the LDOS peaks. Still, if values of $Q_{0}$
are moderate, the LDOS profiles remain qualitatively the same as for
$Q_{0}=0$, and still can be described in terms of the formation of
an antibonding molecular state, as it is illustrated in the panel
(b). However, if the value of the parameter $Q_{0}$ becomes sufficiently
large, the profile of the LDOS dramatically changes. It becomes depleted
in the broad region around the impurities, and corresponds to a distorted
centrosymmetric configuration characteristic to a frustrated atomic
state, as it is shown in the panel (c).

\begin{figure}
\centering{}\includegraphics[width=0.5\textwidth]{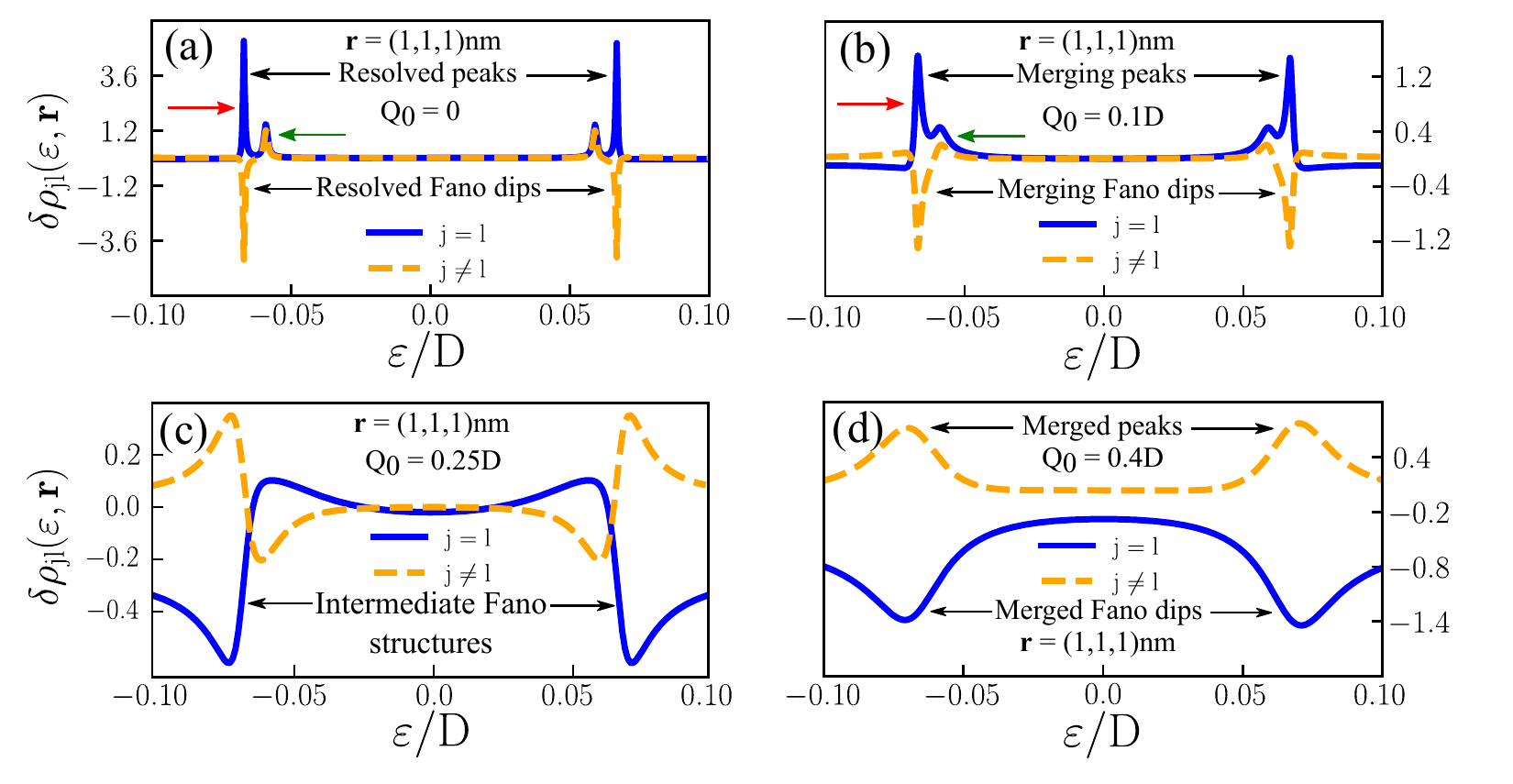} \caption{\label{fig:Pic3} (Color online) Impurity-induced contributions to
the density of states $\delta\rho_{jl}$ as a function of energy.
Position of the STM tip is fixed at $\mathbf{r}=(1,1,1)\:\text{{nm}}$.
Panel (a): The case of a Dirac semimetal host, $Q_{0}=0$. One clearly
sees two well resolved pairs of peaks in $\delta\rho_{jj}$, centered
around $\varepsilon_{d}$ and $\varepsilon_{d}+U$ and corresponding
to bonding (indicated by green arrow) and antibonding (indicated by
red arrow) molecular orbitals. Panel (b): The case of a Weyl metal
host with small value of $Q_{0}=0.1D$. The peaks corresponding to
the molecular states become broadened, but are still clearly resolved.
Panel (c): The case of a Weyl metal host with moderate value of $Q_{0}=0.25D$.
Intermediate Fano structures with merged peaks and dips appear. Panel
(d): The case of a Weyl metal host with large value of $Q_{0}=0.4D$.
Broad plateau in the density of states flanked by a pair of the merged
peaks or dips is formed around $\varepsilon=0$. Transition to the
regime of atomic frustrated state occurs, as seen in Fig.~\ref{fig:Pic2}(c). }
\end{figure}

\begin{figure}
\centering{}\includegraphics[width=0.5\textwidth]{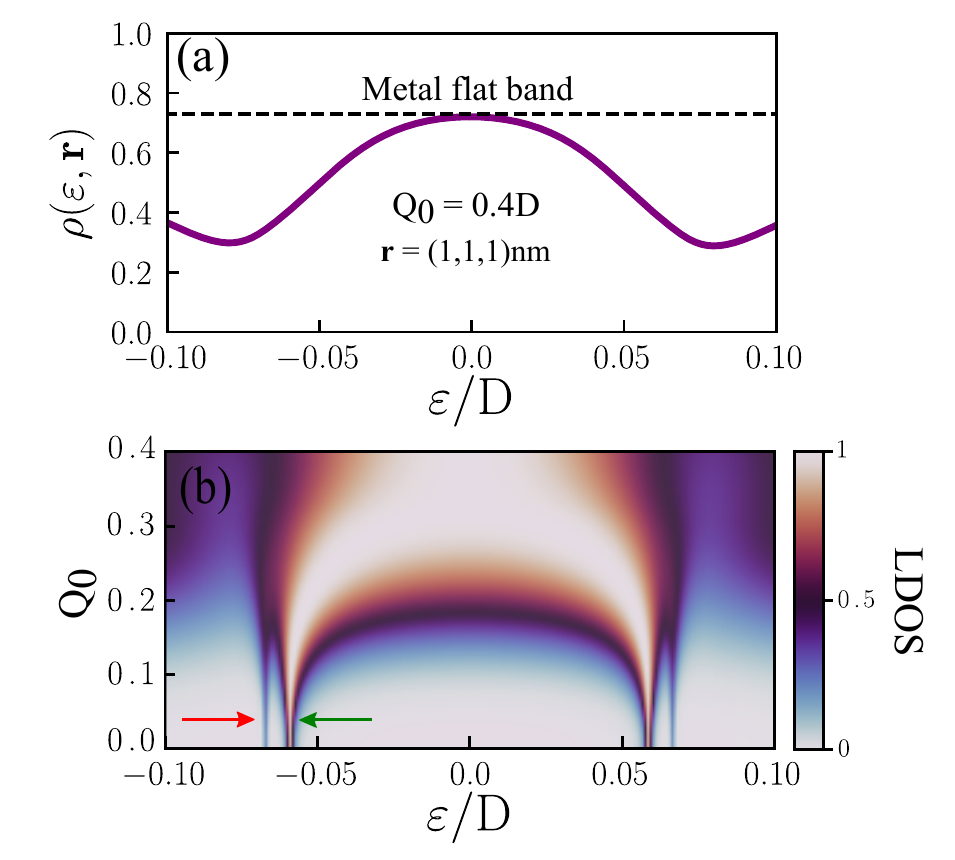} \caption{\label{fig:Pic5}(Color online) Panel (a): The total LDOS of the system
consisting of two impurities placed inside a Weyl metal host with
$Q_{0}=0.4D$, corresponding to the regime of the formation of an
atomic frustrated state. Position of the STM tip is fixed at $\mathbf{r}=(1,1,1)\:\text{{nm}}$.
Panel (b): Phase diagram, showing the total density of sates as function
of the energy $\varepsilon$ and the parameter $Q_{0}$. With increase
of $Q_{0}$ one clearly observes the crossover from the regime of
standard bonding (indicated by green arrow) and antibonding (indicated
by red arrow) molecular orbitals, characterized by four well resolved
Hubbard bands, to the regime of frustrated atomic state.}
\end{figure}

To shed more light on the underlying mechanisms of its formation,
we have analyzed separately different contributions to the LDOS induced
by the impurities, as illustrated by Figs.~\ref{fig:Pic3} and \ref{fig:Pic5}.

\begin{figure}
\centering{}\includegraphics[width=0.48\textwidth]{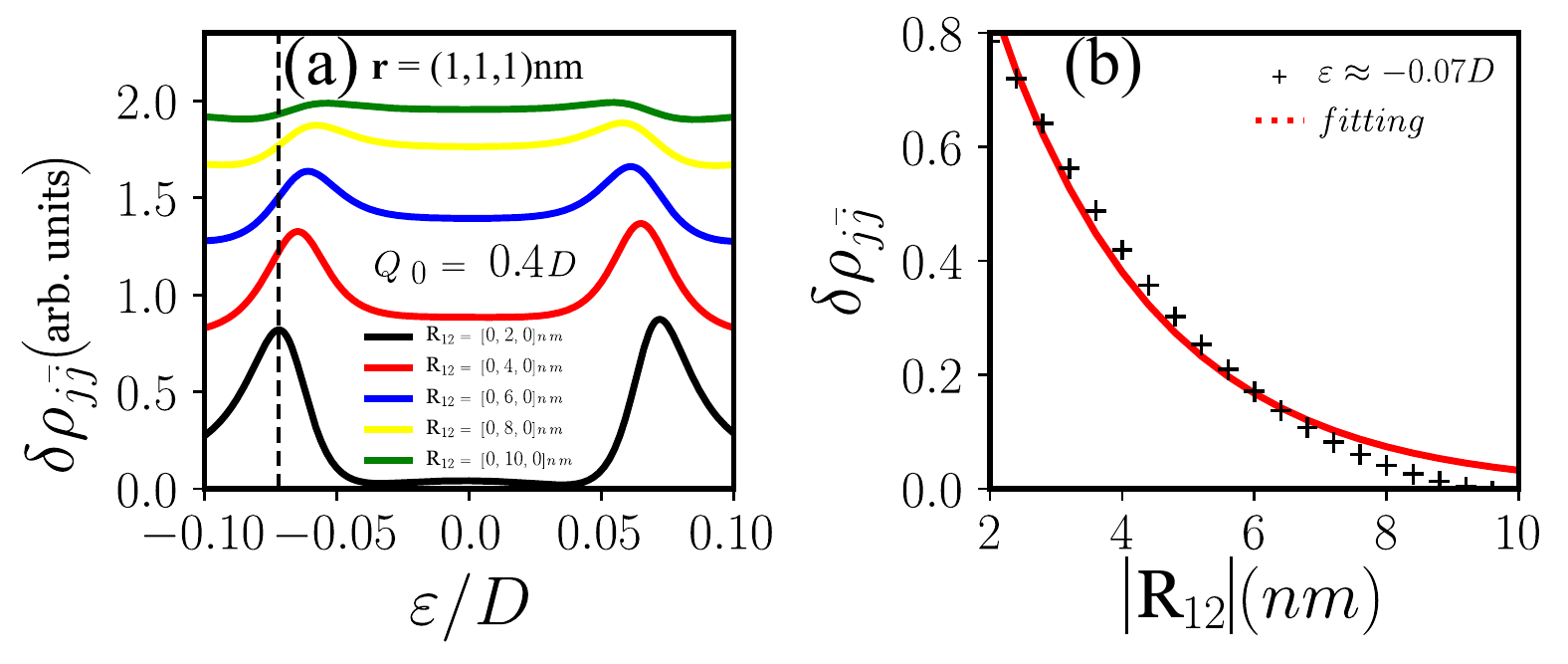} \caption{\label{fig:Pic4}(Color online) Panel (a): Induced LDOS term $\delta\rho_{j\bar{j}}$
($j=1,\bar{j}=2$ and $j=2,\bar{j}=1$) for $Q_{0}=0.4D$ with STM
tip at $\mathbf{r}=(1,1,1)\:\text{{nm}}$ as a function of energy
for several values of ${\bold R_{12}}.$ For a sake of clarity, we
present each case vertically shifted, thus making explicit that the
increasing of ${\bold R_{12}}$ leads to the LDOS $\delta\rho_{j\bar{j}}$
vanishing. Hence, such a quenching reveals the crossover from the
profile with two Hubbard bands, characteristic of the atomic frustrated
state, towards that completely flat, for the uncorrelated pair of
atoms situation. Panel (b): Amplitude of $\delta\rho_{j\bar{j}}$
evaluated at the black-dashed line cut $\varepsilon\approx-0.07D$
marked in panel (a) as a function of $|{\bold R_{12}}|,$ which exhibits
an exponential-like decay (crossed-points in black). Particularly,
it is fitted by $\delta\rho_{j\bar{j}}(\varepsilon\approx-0.07D)=1.96\exp(-0.41|{\bold R_{12}}|)$
(red line).}
\end{figure}

Fig.~\ref{fig:Pic3} shows the plots of $\delta\rho_{jl}$ as a function
of the energy for one particular tip position $\mathbf{r}=(1,1,1)\:\text{{nm}}$
(the change of this latter does not affect the results qualitatively).
Both contributions from intra-impurity ($j=l$) and inter-impurity
($j\neq l$) are shown. In panel (a), corresponding to the case of
a Dirac host with $Q_{0}=0$, one clearly sees the presence of the
four peaks in $\delta\rho_{jj}$, corresponding to well resolved Hubbard
bands and describing the formation of bonding and antibonding molecular
orbitals, which stem from single-impurity bands centered around $\varepsilon_{d}<0$
and $\varepsilon_{d}+U>0$. For the considered parameters, the lowest
energy peak corresponds to the antibonding state (pointed by the red
arrow) and next peak to the bonding molecular state (pointed by the
green arrow) \cite{Marques2017}. The crossed term $\delta\rho_{jl}$,
with $j\neq l$ exhibits two resolved pairs of peaks and Fano dips
instead. The increase of the parameter $Q_{0}$ leads to the broadening
of the peaks and Fano dips (panel (b), $Q_{0}=0.1D$). At some point,
the peaks corresponding to the bonding and antibonding states merge,
giving rise to intermediate Fano lineshapes, with shallow minimum
at $\varepsilon=0$ (panel (c), $Q_{0}=0.25D$). Further increase
of $Q_{0}$ leads to the formation of a broad plateau in the density
of states around $\varepsilon=0$, flanked by a pair of merged peaks
for $j\neq l$, or merged dips for $j=l$ (panel (d), $Q_{0}=0.4D$).
The presence of only two resolved Hubbard bands is typical for a pair
of uncorrelated impurities. However, in our case the amplitudes $\delta\rho_{jl}\neq0$
for $j\neq l$, which means that molecular binding still persists,
although in the unusual form of an atomic frustrated state. In this
configuration, the role of the constructive and destructive Fano interference
channels between $\delta\rho_{jj}$ and $\delta\rho_{jl}$ becomes
inverted with respect to those observed in Dirac hosts, as it can
be clearly seen from the comparison between panels (d) and (a).

We highlight that Figs.~\ref{fig:Pic2} and \ref{fig:Pic3} introduce
the concept of the atomic frustrated state, whose origin is genuinely
of molecular-type, wherein its signatures resemble simultaneously
those from uncorrelated and correlated atoms. Noteworthy, in such
a scenario, the collective behavior of a diatomic molecule mimics
an uncorrelated pair of atoms. Its characterization consists of electronic
depletions (Fano dips) in $\delta\rho_{jj}$ around the Hubbard bands
at $\varepsilon_{d}<0$ and $\varepsilon_{d}+U>0$, exactly as in
the corresponding uncorrelated situation. Additionally and counterintuitively,
a finite inter-impurities correlation ($\delta\rho_{jl}\neq0$ with
$j\neq l$) emerges as in a molecule, being identified by two Hubbard
peaks instead. Hence, as this pair of atoms remains correlated through
the host ($\delta\rho_{jl}$ finite, although with just two Hubbard
structures), but shows itself seemingly uncorrelated ($\delta\rho_{jj}$
with two Hubbard structures, instead of four as in a molecule), then
the state is considered atomically frustrated.

It is worth mentioning that as we focus on the paramagnetic case of
impurities ($\bigl\langle n_{j\uparrow}\bigr\rangle=\bigl\langle n_{j\downarrow}\bigr\rangle),$
the Dzyaloshinskii-Moriya interaction (DMI), which is a type of spin
texture within the RKKY interaction \cite{Chang2015,Hosseini2015}
for Dirac-Weyl semimetals, indeed does not rule this peculiar molecular
binding here reported. The key responsible mechanism for the proposed
state relies on the well-known Friedel-like oscillations \cite{Friedel},
which according to some of us \cite{Marques2017,Marques2019}, by
working cooperatively with the intra-impurities Coulomb repulsion,
is capable of establishing molecular bonds in Dirac-Weyl hosts. Thus,
in the frustration regime of the Weyl metal phase, for first ever,
novel Friedel-like behavior is revealed. It consists of electronic
waves that travel forth and back between the left and right impurities
($\delta\rho_{jl}$), which are entirely phase shifted by $\pi$ with
respect to those scattered locally by the impurities ($\delta\rho_{jj}$).
As a result, $\delta\rho_{jj}$ shows two structures dominantly Fano
destructive at the two Hubbard bands $\varepsilon_{d}<0$ and $\varepsilon_{d}+U>0$
(Fig.~\ref{fig:Pic3}(d)), which then flank a flat metallic-type
plateau in the LDOS around the Fermi energy, as we will see later
on.

In order to understand the long-range behavior of the frustrated atomic
state encoded by $\delta\rho_{jl},$ in Fig.\ref{fig:Pic4}(a) we
analyze such a quantity for $Q_{0}=0.4D$ and STM-tip at $\mathbf{r}=(1,1,1)\:\text{{nm}}$,
as a function of energy upon varying ${\bold R_{12}}.$ We can clearly
perceive that the pair of Hubbard bands of the atomic frustrated state
become broader as we increase the inter-impurities separation, in
such a way that the $\delta\rho_{jl}$ approaches a profile entirely
flat, which corresponds to the case of decoupled atoms. Thus by fixing
the energy, for instance at $\varepsilon\approx-0.07D$ (black-dashed
line cut), it is possible to estimate how quickly $\delta\rho_{jl}$
vanishes. Fig.\ref{fig:Pic4}(b) then makes explicit that the decay
obeys an exponential-like behavior (crossed-points in black), which
is fitted by $\delta\rho_{j\bar{j}}(\varepsilon\approx-0.07D)=1.96\exp(-0.41|{\bold R_{12}}|)$
(red line), where $j=1,\bar{j}=2$ and $j=2,\bar{j}=1$. Notice that
for $|{\bold R_{12}}|=10\:\text{{nm}},$ the molecular bond of the
atomic frustrated state is practically dissociated.

Fig.~\ref{fig:Pic5}(a) shows that the corresponding total LDOS has
very broad maximum at $\varepsilon=0$ and a pair of the broad minima
around $\varepsilon_{d}$ and $\varepsilon_{d}+U.$ We clarify that
the emergence of a flat LDOS (broad plateau) in the vicinity of $\varepsilon=0$
(the Fermi energy) is characteristic of the metallic regime of the
host. This is the direct outcome of the pseudogap closing in Weyl
materials with large $Q_{0}$, for which the host DOS is enhanced
at the Fermi energy. In equivalent words, it leads to the enhancement
of states in the Fermi surface, exactly when the inversion symmetry
is highly broken, which drives the system into the metallic Weyl regime.
This comes from the increasing of the Dirac cones separation obeying
blue and red shifts in the energy axis (see Fig.\ref{fig:Pic1}(c)
of the system sketch). This causes the formation of two bands. However,
distinctly from the spin-orbit coupling, which resolves these bands
in the spin channels, in Weyl metals such a separation occurs in the
chirality degrees of freedom. The latter, we call particular attention,
is defined as the spin projection over the linear momentum, contrasting
the spin-orbit coupling, which instead, projects the former on the
angular momentum. Thus, by integrating all momentum states split in
energy, leads to two branches in the background DOS $\rho_{0}(\varepsilon),$
which are resolved in the chirality degree. Consequently, the branch
with positive chirality shows blue-shift, while the negative presents
the corresponding red (see also Fig.\ref{fig:Pic1}(e)). Such a behavior
closes the pseudogap at the Fermi energy, thus increasing the amount
of the states in the Fermi surface. Therefore, this gives rise to
the flat metallic-type plateau in the LDOS around $\varepsilon=0.$

The crossover between the cases of the standard molecular bonding
and antibonding states, and formation of an atomic frustrated state
is illustrated by Fig.~\ref{fig:Pic5}(b), where a phase diagram,
showing the total LDOS as function of the energy $\varepsilon$ and
the parameter $Q_{0}$ is presented. With increase of $Q_{0}$ the
narrow peaks characteristic to four well resolved Hubbard bands become
broadened and finally merge, producing characteristic profile plotted
in Fig.~\ref{fig:Pic5}(a). From the experimental perspective, such
transition can be achieved by application of stress, which is expected
to break the inversion symmetry~\cite{Armitage2018}.

\section{Conclusions }

We have demonstrated that the nature of electronic states of a pair
of impurities placed inside a Weyl metal strongly depends on the parameter
$Q_{0}$, which defines the breaking of the inversion symmetry in
the host material. For small values of this parameter one observes
the formation of conventional bonding and antibonding molecular orbitals.
However, for large values of $Q_{0}$ transition to an atomic frustrated
state, characterized by a broad bowl-shape distribution of the LDOS
in the real space occurs. This transition should take place under
the application of external stress, which allows to propose the concept
of a molecular switcher, alternating between ordinary molecular and
atomic frustrated states.

\section{Acknowledgments}

We thank the Brazilian funding agencies CNPq (Grants.~305668/2018-8
and 302498/2017-6), the São Paulo Research Foundation (FAPESP; Grant
No. 2018/09413-0) and Coordenação de Aperfeiçoamento de Pessoal de
Nível Superior - Brasil (CAPES) -- Finance Code 001. YM and IAS acknowledge
support the Ministry of Science and Higher Education of Russian Federation,
goszadanie no. 2019-1246, and ITMO 5-100 Program.


\begin{thebibliography}{1}
\bibitem{Kondo2}{The Hubbard I is applicable for temperatures $T\gg T_{K}$,
being $T_{K}$ the Kondo temperature. For the evaluation of $\bigl\langle n_{j\bar{\sigma}}\bigr\rangle,$
$T$ should not be very high so that we can safely assume the Heaviside
step function for the Fermi-Dirac distribution $n_{_{F}}(\varepsilon).$}
\end{thebibliography}
\end{document}